\documentstyle[preprint,aps]{revtex}
\input{epsf.sty}
\begin{document}
\draft
\title{Gluon and gluino penguins and the charmless decays of the $b$ quark}
\author{S A Abel}
\address{ Theory Division, CERN CH-1211, Geneva 23, Switzerland}
\author{W N Cottingham}
\address{H H Wills Physics Laboratory,\\ 
Royal Fort, Tyndall Ave, Bristol, BS8 1TL, UK}
\author{I B Whittingham}
\address{School of Computer Science, Mathematics and Physics,\\ 
James Cook University, Townsville, Australia, 4811}
\date{\today}
\maketitle
\title{Abstract}
\begin{abstract}
Gluon mediated exclusive hadronic decays of $b$ quarks  are studied within
the standard model (SM) and the constrained minimally supersymmetric
standard model (MSSM). For all allowed regions of the MSSM parameter space
$(A, \;\tan \beta ,\;m_{0},\;m_{1/2})$ the penguin magnetic 
dipole form factor $F^{R}_{2}$ is dominant over the electric dipole and
can be larger than the magnetic dipole form factor of the SM. However,
overall the SM electric dipole decay amplitude $F^{L}_{1}$ dominates 
the decay rate.
The MSSM penguin contributions to the free quark decay rate approach the 
10\% level for those regions of parameter space close to the highest allowed 
values of $\tan \beta $ $(\sim 55)$ for which the gluino is light 
$(m_{\tilde{g}} \approx 360 $ GeV) and lies within the range of the 
six $\tilde{d}$ squark masses. In these regions the supersymmetric box 
amplitudes are negligible. The MSSM phases change very little over the
allowed parameter space and can lead to significant interference 
with the SM amplitudes.
\end{abstract}
\narrowtext
\section{Introduction}

Supersymmetry (SUSY) is a highly favoured candidate theory for new physics
beyond the standard model (SM). Of particular interest are the flavour
changing neutral current transitions involving the quark-squark-gluino 
vertex and the non-removable CP-violating phases which arise as the 
renormalisation group equations (RGE) scale the physics down from the 
unification scale $M_{U} \sim 10^{16}$ GeV to the electroweak scale. These
effects of SUSY have implications for rare B decays and mixings 
\cite{Bert87,Bert91,Bigi91} and for other observables such as 
quark electric dipole moments \cite{edm,Abel96}. 

Measurements of rare flavour-changing B decays provide opportunities for
the discovery of indirect effects of SUSY \cite{Gronau97,Hewett97} as the
measured observables involve SM and SUSY processes occurring at the same
order of perturbation theory. In contrast to the situation for 
$B^{0}-\bar{B}^{0}$ mixing where new physics is expected to change the 
magnitude of the CP-asymmetries but not the patterns of asymmetries 
predicted by the SM \cite{Gross97}, the effects of new physics in decay
amplitudes depends on the specific processes and decay channel under
consideration and, although small, may be detectable by comparing 
measurements that within the SM should yield the same quantity.

The $b \rightarrow s$ transition provides an opportunity to study CP 
violation from non-standard phases \cite{Atwood97} and there is significant
current interest in the $b \rightarrow sg$ penguin decay for which it has 
been argued \cite{Hou97} that enhancement for on-shell gluons is needed
from non-SM physics to explain the CLEO measurement \cite{CLEO} of a large
branching ratio for $B \rightarrow \eta^{\prime}+X_{s}$ and the 
$\eta^{\prime}-g-g$ gluon anomaly.

For the gluon-mediated exclusive hadronic decays studied here the effects
of SUSY are expected to be small and difficult to disentangle from the SM
effects because of the large uncertainties associated with the SM 
predictions. The SM calculations involve \cite{Gronau97} the computation
of the quark level decays $b \rightarrow qq^{\prime}\bar{q}^{\prime}$,
calculation of the Wilson coefficients \cite{Buch96} to incorporate QCD
corrections as the physics is renormalised down from the electroweak scale
to the scale $m_{b}$ and, finally, the calculation of hadronic matrix 
elements for the hadronisation of the final-state quarks into particular
final states, typically evaluated using the factorisation assumption. As
this last stage can introduce such large uncertainties that predicted SM
rates for exclusive hadronic penguin decays can be in error by a factor 
of 2 to 3, we will restrict the present study to the weak scale quark 
level processes where any differences between SM and SUSY physics are more
apparent.

The most predictive of the SUSY models is the (constrained) minimally 
supersymmetric standard model (MSSM) \cite{MSSM,Abel96} based on 
spontaneously broken $N=1$ supergravity with flat K\"{a}hler metrics 
\cite{SUGRA}, universal explicit soft-SUSY breaking terms at the scale
$M_{{\mathrm MSSM}}\sim M_{U}$ and spontaneous breaking of the 
$SU(2)\otimes U(1)$ symmetry driven by radiative corrections. Such models
contain two CP-violating phases $\delta^{{\mathrm MSSM}}_{A,B}$ from the
soft-SUSY breaking terms in addition to the usual phase 
$\delta_{{\mathrm CKM}}$ of the CKM mixing matrix. With the usual 
assumption that these SUSY phases vanish identically at the unification
scale because of CP conservation in the SUSY breaking sector, it is claimed
\cite{Ciuch97,Gronau97,Bert91,Bigi91} that the MSSM predictions for 
$B^{0}-\bar{B}^{0}$ mixing and penguin decays such as $b \rightarrow q
q^{\prime}\bar{q}^{\prime}$ are very similar to those of the SM and that
non-minimal SUSY models are needed to obtain any significant  non-SM effects.
An early study \cite{Lang84} concluded that superpenguins are small
compared to ordinary penguins unless the gluino is very light ($\approx 1 $
GeV) and satisfies $m_{\tilde{g}} \ll m_{\tilde{d}}$.
Recently Grossman and Worah \cite{Gross97} have found that the gluonic penguin
amplitudes for $b \rightarrow sq\bar{q}$ and $b \rightarrow dq\bar{q}$ in 
the effective SUSY model of Cohen {\it et al} \cite{Cohen97} can be up to
twice as large as the SM gluonic penguins and with an unknown phase.

In this paper we revisit the question of MSSM predictions for the penguin
mediated decays $b \rightarrow qq^{\prime}\bar{q}^{\prime}$. In doing
so we review in some detail the SM predictions with particular reference
to the relative contributions of the internal $u$, $c$ and $t$ quarks 
to the  gluon penguin \cite{Fleisch97},the relative magnitudes of the various
form factors and the role of the strong and weak phases \cite{Band79,Gerard91}.
We find, for example, that the CP violating phases for the $b \rightarrow 
d g$ and $b \rightarrow s g $ electric form factors, which dominate the
decay amplitude, have no simple relationship with any angle of the 
unitarity triangle. For the MSSM we explore the allowed regions of the 
parameter space to locate those regions which give the largest 
modifications to the SM results. In contrast to the SM, we always find
the magnetic amplitude to dominate the electric amplitude. Also, there
are large regions of the MSSM parameter space for which the magnetic 
amplitude is greater than that of the SM. The search for SUSY would be 
greatly aided if the magnetic amplitudes could be experimentally isolated.

Conservation of the gluonic current requires the $b \rightarrow q g$ vertex
to have the structure
\begin{equation}
\label{b1}
\Gamma^{a}_{\mu}(q^{2})=\frac{ig_{s}}{4 \pi ^{2}}\;{\bar u}_{q}(p_{q})T^{a}
V_{\mu}(q^{2})u_{b}(p_{b})
\end{equation}
where
\begin{eqnarray}
\label{b2}
V_{\mu}(q^{2})&=&(q^{2}g_{\mu\nu}-q_{\mu}q_{\nu})\gamma^{\nu}[F^{L}_{1}
(q^{2})P_{L} + F^{R}_{1}(q^{2})P_{R}] \nonumber  \\
&&+ i \sigma_{\mu\nu}q^{\nu}[F^{L}_{2}(q^{2})
P_{L} + F^{R}_{2}(q^{2})P_{R}].
\end{eqnarray}
Here $F_{1}$ and $F_{2}$ are the electric (monopole) and magnetic (dipole)
form factors, 
$q=p_{b}-p_{q}$ is the gluon momentum, $P_{L(R)}\equiv (1\mp \gamma_{5})/2$
 are the chirality
projection operators and $T^{a}\;(a=1,\ldots,8)$ are the $SU(3)_{c}$ 
generators normalised to $Tr(T^{a}T^{b})=\frac{1}{2}\delta^{ab}$.

The ${\bar b} \rightarrow {\bar q}g$ vertex is
\begin{equation}
\label{bb1}
{\bar \Gamma}^{a}_{\mu}(q^{2})=-\frac{ig_{s}}{4\pi^{2}}\;{\bar v}_{b}(p_{b})
T^{a}{\bar V}_{\mu}(q^{2})v_{q}(p_{q})
\end{equation}
where ${\bar V}_{\mu}$ has the form (\ref{b2}) with the form factors
$F^{L,R}_{1,2}(q^{2})$ replaced by $\bar{F}^{L,R}_{1,2}(q^{2})$ where the
relationship between the $F$ and $\bar{F}$ form factors will be 
discussed later.

To lowest order in $\alpha_{s}$ the penguin amplitude for the 
decay process $b \rightarrow q\;g 
\rightarrow q\;q^{\prime}\bar{q}^{\prime}$ is
\begin{equation}
\label{b46}
M^{\text{Peng}}=-\frac{ig^{2}_{s}}{4 \pi^{2}}\;
[\bar{u}_{q}(p_{q})T^{a}\hat{\gamma}_{\mu}u_{b}(p_{b})]
[\bar{u}_{q^{\prime}}(p_{q^{\prime}})
\gamma^{\mu}T^{a}v_{\bar{q}^{\prime}}(p_{\bar{q}^{\prime}})]
\end{equation}
where
\begin{eqnarray}
\label{b47}
\hat{\gamma}_{\mu}&\equiv &\gamma_{\mu}[F^{L}_{1}(q^{2})P_{L}+
F^{R}_{1}(q^{2})P_{R}] \nonumber  \\
&&+\frac{i\sigma_{\mu\nu}q^{\nu}}{q^{2}}[F^{L}_{2}
(q^{2})P_{L}+F^{R}_{2}(q^{2})P_{R}].
\end{eqnarray}
This gives the free quark decay rate
\begin{equation}
\label{b48}
\frac{d\Gamma^{\text{Peng}}}{dq^{2}}=\frac{1}{288\pi^{3}}(\frac{g_{s}^{2}}
{4\pi^{2}})^{2}
\frac{1}{E_{b}}I(q^{2})(1+\frac{2m^{2}_{q^{\prime}}}{q^{2}})\;N(q^{2})
\end{equation}
where
\begin{equation}
\label{b49}
I(q^{2})=[1+\frac{(m^{2}_{q}-q^{2})^{2}}{m_{b}^{4}}-2\frac{m^{2}_{q}+q^{2}}
{m_{b}^{2}}]^{1/2}\;\;[1-\frac{4m^{2}_{q^{\prime}}}{q^{2}}]^{1/2}
\end{equation}
is the phase space factor and
\widetext
\begin{eqnarray}
\label{b50}
N(q^{2})&=&(q^{2}\;p_{b}\cdot p_{q}+2\;p_{b}\cdot q\; p_{q}\cdot q)
(|F^{L}_{1}|^{2}+|F^{R}_{1}|^{2})-3m_{b}m_{q}q^{2}(F^{L}_{1}F^{R\;*}_{1} 
+ c.c.) \nonumber  \\
&& -3m_{b}\;p_{q}\cdot q (F^{L}_{1}F^{R\;*}_{2}+F^{R}_{1}F^{L\;*}_{2} + c. c.)
        +3m_{q}\;p_{b}\cdot q(F^{L}_{1}F^{L\;*}_{2}+F^{R}_{1}F^{R\;*}_{2}
        + c. c.)  \nonumber  \\
&& +\frac{1}{q^{2}}(4\;p_{b}\cdot q\; p_{q}\cdot q - q^{2}\;p_{b}\cdot p_{q})
        (|F^{L}_{2}|^{2}+|F^{R}_{2}|^{2}) 
        -3 m_{b}m_{q}(F^{L}_{2}F^{R\;*}_{2} + c. c.)
\end{eqnarray}
\narrowtext
with $4m_{q^{\prime}}^{2} \leq q^{2} \leq (m_{b}-m_{q})^{2}$. 

Similarly, for ${\bar b}\rightarrow {\bar q}\;q^{\prime}\;{\bar q}^{\prime}$,
the amplitude is
\begin{equation}
\label{bb2}
{\bar M}^{\text{Peng}}=\frac{ig_{s}^{2}}{4 \pi^{2}}\;
[\bar{v}_{q}(p_{q})T^{a}\bar{\hat{\gamma}}_{\mu}v_{b}(p_{b})]
\bar{u}_{q^{\prime}}(p_{q^{\prime}})
\gamma^{\mu}T^{a}v_{\bar{q}^{\prime}}(p_{\bar{q}^{\prime}})]
\end{equation}
where $\bar{\hat{\gamma_{\mu}}}$ is obtained from (\ref{b47}) by the 
replacement of all $F(q^{2})$ form factors by $\bar{F}(q^{2})$ form factors. 
The decay rate $d\bar{\Gamma}^{\text{Peng}}/dq^{2}$ 
is given by (\ref{b48})--(\ref{b50}) with the same replacements.

One CP violation observable of particular interest is the partial rate 
asymmetry
\begin{equation}
\label{bb3}
{\cal A}_{\text{CP}}(q^{2}) \equiv \frac{d\Gamma_{A}/dq^{2}}
{d\Gamma_{S}/dq^{2}}
\end{equation}
where
\begin{equation}
\label{bb4}
\frac{d\Gamma_{S/A}}{dq^{2}}=\frac{1}{2}\;(\frac{d\Gamma}{dq^{2}}\pm
\frac{d\bar{\Gamma}}{dq^{2}}).
\end{equation}

\section{The gluon penguin in the Standard Model}

Although the expressions for the SM form factors are known, we believe it is
useful to present a brief outline of their derivation, both in order to 
clarify their regimes of validity and to aid our later generalisation to 
include the effects of SUSY.
 
\begin{figure}
  \begin{center}
    \leavevmode   
    \hbox{\epsfxsize=0.8\textwidth
      \epsffile{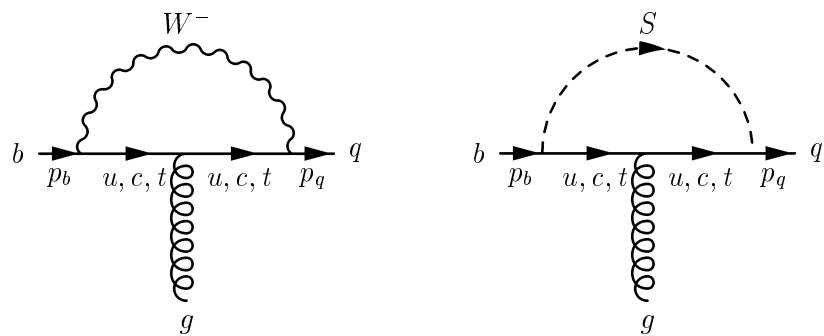}}
  \end{center}
  \caption{SM gluon penguins with $W$ and scalar exchange.}
  \label{fig1} 
\end{figure}

For the SM the contributions to the $b \rightarrow qg$ vertex 
$\Gamma^{a}_{\mu}$ from $W$ and scalar exchange (Fig. \ref{fig1}) give 
\cite{Desh82,Desh83}
\begin{equation}
\label{b3}
V^{\text{SM}}_{\mu}(q^{2})= \sum_{i=u,c,t}\;\lambda^{bq}_{i}[(A^{W}_{\mu}
+ A^{S}_{\mu})P_{L} + B^{S}_{\mu}P_{R}]
\end{equation}
where $\lambda^{bq}_{i}\equiv \frac{g_{2}^{2}}{8M_{W}^{2}}\;
K_{iq}^{*}K_{ib}$ and $K$ is the CKM matrix. For the $\bar{b}
\rightarrow \bar{q}g$ vertex $\bar{V}_{\mu}$, the CKM matrix
elements are replaced by their complex conjugates.

After putting the external
$b$ and $q\;(=d,s)$ quarks on mass shell, $V^{\text{SM}}_{\mu}$ has the
form
\begin{eqnarray}
\label{b4}
V^{\text{SM}}_{\mu}(q^{2})&=&(a\; p_{b\mu} + b\; q_{\mu} + c\; 
\gamma_{\mu})P_{L} \nonumber  \\
&&+ (d\; p_{b\mu} +e\; q_{\mu} + f\; \gamma_{\mu})P_{R}
\end{eqnarray}
and the form factors for arbitrary $q^{2}$ are given by
\begin{eqnarray}
\label{b5}
2(m_{q}^{2}-m_{b}^{2})F^{L}_{1}(q^{2})& = & (a + 2b)m_{q} + (d+2e) m_{b} \\
\label{b6}
2(m_{q}^{2}-m_{b}^{2})F^{R}_{1}(q^{2})& = &  (a +2b) m_{b} + (d+2e) m_{q}
\end{eqnarray}
and
\begin{equation}
\label{b7}
F^{L}_{2}(q^{2})=a/2, \quad F^{R}_{2}(q^{2})=d/2.
\end{equation}

Neglecting terms of order $m_{q}^{2}/M_{W}^{2} $ and $m_{b}^{2}/M_{W}^{2}$
then
\begin{equation}
\label{b8}
a + 2b =-m_{q}\;\alpha^{\text{SM}},\quad d +2e = m_{b}\;\beta^{\text{SM}}
\end{equation}
where
\begin{eqnarray}
\label{b9}
\alpha^{\text{SM}}(q^{2})&=&\sum_{i}\;\lambda^{bq}_{i}\int^{1}_{0} dx
\int^{1-x}_{0} dy [4(x+2y)(y-1)  \nonumber  \\
                &&-2x_{i}(1-x-xy-2y^{2})]/Y_{i}(x,y)  \\
\label{b10}
\beta^{\text{SM}}(q^{2})&=&\sum_{i}\;\lambda^{bq}_{i}\int^{1}_{0} dx
\int^{1-x}_{0} dy [4x^{2}-8x \nonumber  \\
&&+8y^{2}-8y+12xy +x_{i}(4x^{2}-6x+4y^{2} \nonumber  \\
&& -8y+10xy+2)]/Y_{i}(x,y)
\end{eqnarray}
together with
\begin{eqnarray}
\label{b11}
F^{L}_{2}(q^{2})&=&m_{q}\;\sum_{i}\;\lambda^{bq}_{i}\int^{1}_{0} dx 
\int^{1-x}_{0} dy [2x(1-y)  \nonumber  \\
&&+x_{i}(1-x-xy)]/Y_{i}(x,y)
\end{eqnarray}
and
\begin{eqnarray}
\label{b12}
F^{R}_{2}(q^{2})&=&m_{b}\;\sum_{i}\;\lambda^{bq}_{i}\int^{1}_{0} dx 
\int^{1-x}_{0} dy [2x(x+y) \nonumber  \\
&&+x_{i}(2x^{2}-3x +3xy +1)]/Y_{i}(x,y).
\end{eqnarray}
In the above
\begin{equation}
\label{b13}
Y_{i}(x,y)=x+x_{i}(1-x)+q^{2}[xy+y(y-1)]/M_{W}^{2}
\end{equation}
where $x_{i}\equiv m_{i}^{2}/M_{W}^{2}$. 

If these expressions are evaluated at $q^{2}=0$ we have $\beta^{\text{SM}}=
\alpha^{\text{SM}}$ and
\begin{eqnarray}
\label{b14}
F^{L}_{1}(0)&=&\frac{g^{2}_{2}}{8M_{W}^{2}}\sum_{i}\;K^{*}_{iq}K_{ib}
\;f_{1}(x_{i}),\quad F^{R}_{1}(0)=0 \\
\label{b15}
\frac{1}{m_{q}}F^{L}_{2}(0)&=&\frac{1}{m_{b}}F^{R}_{2}(0)=
\frac{g^{2}_{2}}{8M_{W}^{2}}\sum_{i}\;
K^{*}_{iq}K_{ib}\;f_{2}(x_{i})
\end{eqnarray}
where \cite{Inami81,Hou88}
\begin{eqnarray}
\label{b16}
f_{1}(x)&=&\frac{1}{12(1-x)^{4}}[18x-29x^{2}+10x^{3}+x^{4} \nonumber  \\
&& - (8-32x+18x^{2}) \ln x],  \\
f_{2}(x)&=&\frac{-x}{4(1-x)^{4}}[2+3x-6x^{2}+x^{3}+6x \ln x].
\end{eqnarray}
For small $x_{i}$, $f_{2}(x_{i})\approx \frac{1}{2}x_{i}$ whereas 
$f_{1}(x_{i}) \approx -\frac{2}{3} \ln x_{i} $.

For $b\rightarrow q g$, $(q^{2})_{\text{max}}=(m_{b}-m_{q})^{2} \approx
20$ GeV$^{2}$ and the assumption $q^{2}\ll m_{i}^{2}$ which would
justify the replacement of the form factors with their values at 
$q^{2}=0$  is invalid for $F_{1}(q^{2})$ for the $u$ and $c$ quarks. 
This observation has also been made in \cite{Atwood97,Fleisch97}. For these 
light quarks we can evaluate $F^{L}_{1}(q^{2})$ by neglecting $m_{q}^{2}$
compared to $m_{b}^{2}$ in (\ref{b5}) and $x_{i}$ in the numerator of
(\ref{b10}) so that
\begin{eqnarray}
\label{b17}
f_{1}(x_{i},q^{2})&=&-\int^{1}_{0} dx\int^{1-x}_{0} dy 
[2x^{2}-4x+6xy    \nonumber  \\
&&  +4y(y-1)]/Y_{i}(x,y).
\end{eqnarray}
This integral is dominated by the logarithmic singularity near $x=0$ so
we can set $x=0$ everywhere except in the leading term of the denominator
to give
\begin{eqnarray}
\label{b18}
f_{1}(x_{i},q^{2}) &\approx & 4\int^{1}_{0} dx \int^{1}_{0} dy \frac{y(1-y)}
{x+x_{i}-q^{2}y(1-y)/M_{W}^{2}}  \\
\label{b19}
&=&\frac{10}{9} -\frac{2}{3}\ln x_{i} +\frac{2}{3z_{i}} -\frac{2}{3} 
\frac{2z_{i}+1}{z_{i}}\;g(z_{i})
\end{eqnarray}
where $z_{i}\equiv q^{2}/4m_{i}^{2}$ and
\begin{equation}
g(z)=\left\{ \begin{array}{ll}
             \sqrt{\frac{1-z}{z}}\;\arctan (\sqrt{\frac{z}{1-z}}), & z<1 \\
             \frac{1}{2}\sqrt{\frac{z-1}{z}}[\ln (\frac{\sqrt{z}+\sqrt{z-1}}
             {\sqrt{z}-\sqrt{z-1}}) - i \pi],  & z>1
             \end{array}
     \right.
\end{equation}
For $q^{2} > 4 m^{2}_{i}$, $g(z)$ becomes imaginary due to the generation
of a strong phase at the $u\bar{u}$ and $c\bar{c}$ thresholds 
\cite{Band79,Gerard91}.
Our result for $f_{1}(x_{i},q^{2})$ is equivalent to that obtained by
Gerard and Hou \cite{Gerard91}.
For the $u$ quark, $z_{i}$ is large and we use the asymptotic form of 
(\ref{b19}):
\begin{equation}
\label{b20}
f_{1}(x_{u},q^{2}) = \frac{10}{9} -\frac{2}{3}[\ln (\frac{q^{2}}{M_{W}^{2}}) 
-i \pi].
\end{equation}

We will be concerned with the $b \rightarrow d q^{\prime}\bar{q}^{\prime}$ 
and $b \rightarrow s q^{\prime} \bar{q}^{\prime}$ transitions. Although the
form factors $F_{1}$ and $\frac{1}{m_{b}}F_{2}$ contribute to the decay
amplitudes (\ref{b46}) and (\ref{bb2}) with different kinematic factors,
we find that globally over all phase space (but with $q^{2} \geq 1$ GeV$^{2}$)
the kinematic factors are approximately of equal weight which makes it
useful to compare the overall magnitudes of the form factors. We find
$F^{L}_{1} \gg F^{R}_{1}$ and $F^{R}_{2}\gg F^{L}_{2}$. For the 
$b \rightarrow d q^{\prime} \bar{q}^{\prime}$ amplitude we find that
$F^{L}_{1}$ is dominant $(\frac{1}{m_{b}}|F^{R}_{2}| \lesssim \frac{1}{30}
|F^{L}_{1}|)$.

\begin{figure}
  \begin{center}
    \leavevmode   
    \hbox{\epsfxsize=0.8\textwidth
      \epsffile{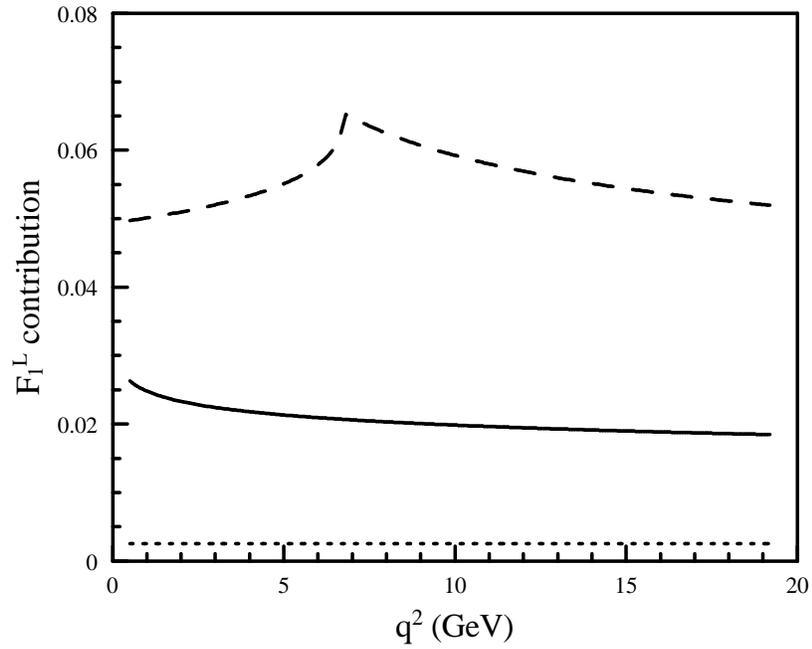}}
  \end{center}
  \caption{Contributions to SM $F^{L}_{1}(q^{2})$ for $ b \rightarrow
   d + g $ from $u$ (solid line), $c$ (dashed line) and $t$ (dotted line)
   quarks.}
  \label{fig2} 
\end{figure}

 The individual contributions $|K^{*}_{id}K_{ib}f_{1}(x_{i},
q^{2})|$, ($i=u,c,t$), to $F^{L}_{1}$ are shown in Fig. \ref{fig2}. 
These magnitudes are the same for 
the $\bar{b}\rightarrow \bar{d}+g$ transition. The $c$ quark is the largest
contributor. The weak phase from the CKM matrix is very small but this 
contribution carries a strong phase for $q^{2} > 4 m^{2}_{c}$. This strong 
phase is the same for the $\bar{b}\rightarrow \bar{d}+g$ transition.
The $u$ quark contribution has a weak CP-violating phase $e^{-i\delta_{13}}
\approx e^{-i\gamma}$ (Particle Data Group notation) and also a 
strong phase that is common
to the $\bar{b}\rightarrow \bar{d}+g$ transition. The $t$ quark contribution
is negligible.

\begin{figure}
  \begin{center}
    \leavevmode   
    \hbox{\epsfxsize=0.8\textwidth
      \epsffile{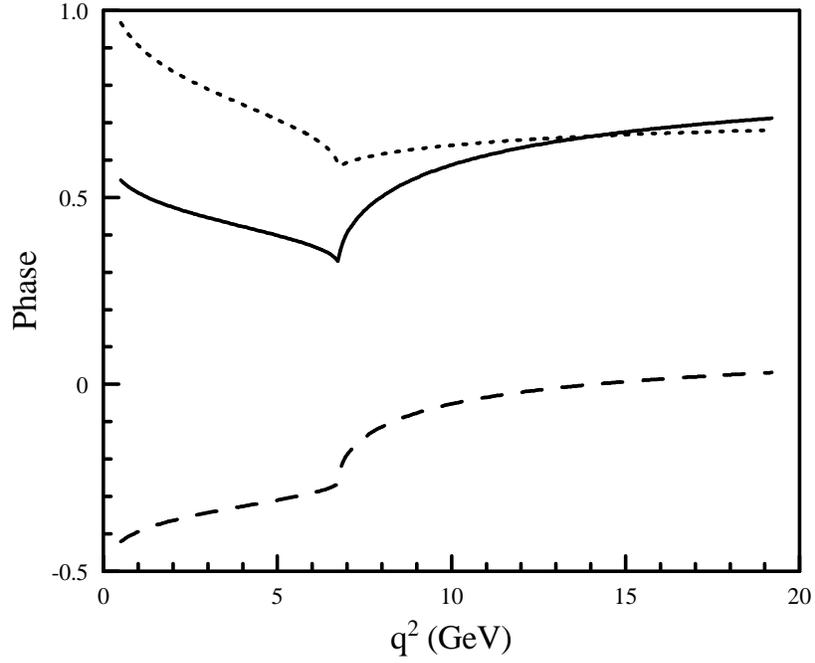}}
  \end{center}
  \caption{Phase of SM $F^{L}_{1}(q^{2})$ for $b \rightarrow d + g$ 
  (solid line) and $\bar{b} \rightarrow \bar{d} + g$ (dashed line) for a 
  CKM phase of $\pi/2$. The dotted line shows the CP-violating phase 
  difference.}
  \label{fig3} 
\end{figure}

These individual amplitudes add to make $F^{L}_{1}$ and, because $u$ 
and $c$ make significant contributions, the phase of $F^{L}_{1}$ differs
for the $b \rightarrow d+g$ and $\bar{b} \rightarrow \bar{d}+g$ transitions.
The phase difference, which can be called the net CP-violating phase, is not
negligible but has no simple relationship with any particular angle of the 
unitary triangle. With $\delta_{13}=\pi/2$ and $s_{13}=0.0035$, we show this
phase in Fig. \ref{fig3}.

\begin{figure}
  \begin{center}
    \leavevmode   
    \hbox{\epsfxsize=0.8\textwidth
      \epsffile{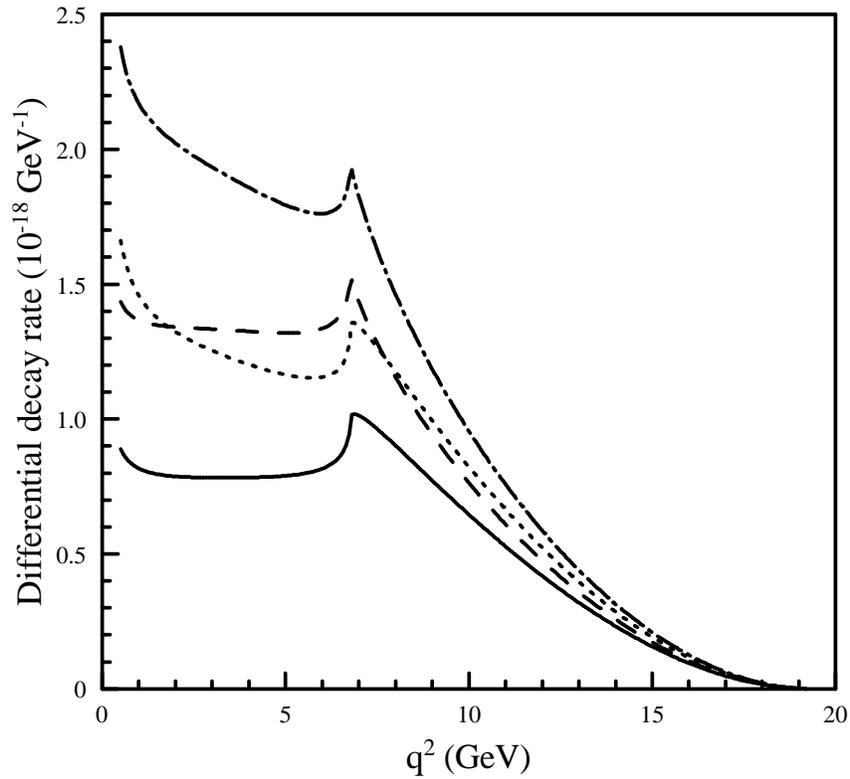}}
  \end{center}
  \caption{Differential SM decay rates for $b \rightarrow d s \bar{s}$
  (solid line) and $ \bar{b} \rightarrow \bar{d} s \bar{s}$ (dashed line). The
  corresponding results for the combined effects of the SM and MSSM are 
  given by the dotted and dot-dashed lines respectively.}
  \label{fig4} 
\end{figure}

Because of the presence of both strong and weak phases the magnitudes of
$F^{L}_{1}$ are also different for the $b$ and $\bar{b}$ decays. 
Processes like $b \rightarrow d s \bar{s}$
and $\bar{b}\rightarrow \bar{d} s \bar{s}$ are expected to be penguin 
dominated and $F^{L}_{1}$ dominates all the other form factors. The decay 
rates $d\Gamma /dq^{2}$ calculated from (\ref{b48}) are shown in 
Fig. \ref{fig4}. The $c\bar{c}$
threshold cusp is clearly exhibited and CP violation is manifest. 
The difference of the decay rates can easily be shown to be proportional
to the Jarlskog factor \cite{Jarls85} $\Im [K_{ub}K^{*}_{ud}K_{cd}
K^{*}_{cb}]=c_{12}c^{2}_{13}c_{23}s_{12}s_{13}s_{23}\sin \delta_{13}$.
Since this factor basically controls the magnitude of the asymmetry, the
modification with different choices of $s_{13}$ and $\delta_{13}$ (the
least known elements of the CKM matrix) can be assessed. The asymmetry is 
large because the sum of the decay rates is also small. 

\begin{figure}
  \begin{center}
    \leavevmode   
    \hbox{\epsfxsize=0.8\textwidth
      \epsffile{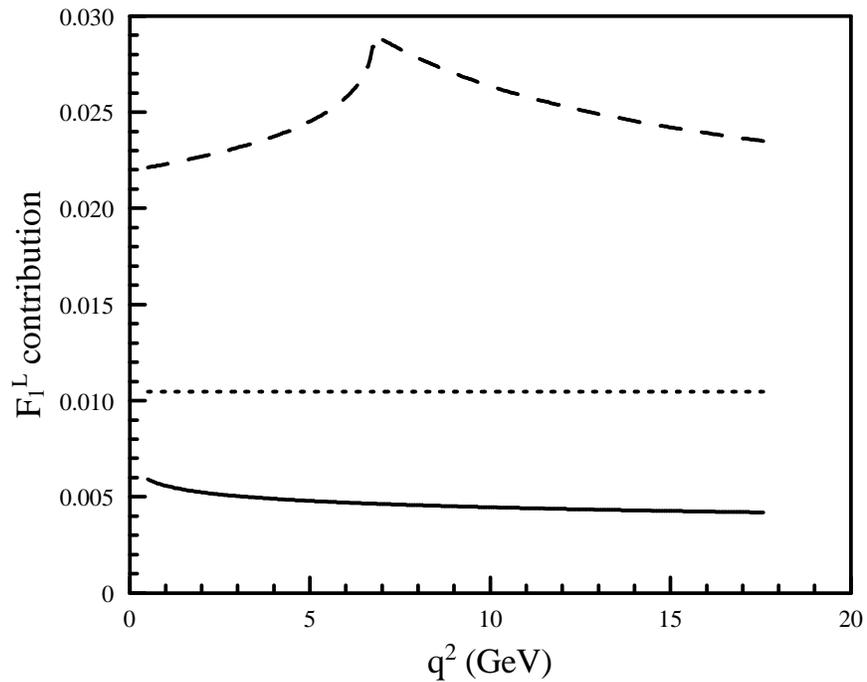}}
  \end{center}
  \caption{Contributions to SM $F^{L}_{1}(q^{2})$ for $ b \rightarrow
  s + g $ from $u$ (solid line), $c$ (dashed line) and $t$ (dotted line)
  quarks.  The contribution from the $c$ quark has been scaled down by a 
  factor of 10.}
  \label{fig5} 
\end{figure}

Turning to the $b \rightarrow s q^{\prime} \bar{q}^{\prime}$ transition, 
we again find that $F^{L}_{1} \gg F^{R}_{1}$, $F^{R}_{2} \gg F^{L}_{2}$ 
and the $F^{L}_{1}$  amplitude to be dominant. 
The individual contributions from $u$, $c$ and $t$
are shown in Fig. \ref{fig5}.  The $c$ quark contribution in this case 
greatly outweighs that of the $u$ and $t$ quarks and since its 
contribution is so large and has almost zero weak phase, the weak
phase on $F^{L}_{1}$ is very small. Processes like 
$b \rightarrow s d \bar{d}$
and $\bar{b} \rightarrow \bar{s} d \bar{d}$ are expected to be penguin 
dominated and these lowest order calculations give the decay rates shown in 
Fig. \ref{fig6}.

\section{The gluino penguin in the MSSM}

In the MSSM there are contributions to $\Gamma^{a}_{\mu}$ from the two 
gluino exchange diagrams I and II (Fig. \ref{fig7}) corresponding to 
the gluon line attached respectively to the gluino and $\tilde{d}$ squark
lines. The MSSM penguin amplitudes 
have the form
\begin{eqnarray}
\label{b21}
V^{\text{MSSM}}_{AB}(q^{2})&=&\sum_{j}\;\Lambda^{bq}_{ABj}\;\{C_{2}(G)
A^{\text{I}}_{AB\mu}    \nonumber  \\
&&+[-C_{2}(G)+2C_{2}(R)]A^{\text{II}}_{AB\mu}\}P_{A}
\end{eqnarray}
where $(A,B)$ are chirality indices, $C_{2}(G)=3$ and $C_{2}(R)=
\sum_{a}T^{a}T^{a}=4/3$ are $SU(3)$ Casimir invariants and $j=1,\ldots,6$ 
labels the $d$ squark mass eigenstates. The coefficient
\begin{equation}
\label{bbb1}
 \Lambda^{bq}_{ABj} \equiv - \frac{g_{s}^{2}}{4m_{\tilde{g}}^{2}}
V^{jq\;*}_{\tilde{d}A}\;V^{jb}_{\tilde{d}B}
\end{equation}
describes the rotation from the down-diagonal 
interaction states to the $\tilde{d}$ mass eigenstates at the 
$d-\tilde{d}-\tilde{g}$ vertices. The matrices $V_{\tilde{d}L}$ and
$V_{\tilde{d}R}$ are obtained from the $(6\times 6)$ matrix $V_{\tilde{d}}=
(V_{\tilde{d}L},V_{\tilde{d}R})^{T}$ which diagonalises the 
$\tilde{d}$ mass$^{2}$ matrix
\begin{equation}
M^{2}_{\tilde{d}}= \left( \begin{array}{ll}
                        (M^{2}_{\tilde{d}})_{LL} & (M^{2}_{\tilde{d}})_{LR} \\
                          &  \\
                        (M^{2}_{\tilde{d}})_{RL} & (M^{2}_{\tilde{d}})_{RR}
                        \end{array}
                  \right).  \nonumber
\end{equation}
Placing the external quarks on mass shell converts 
$V^{\text{MSSM}}_{AB}(q^{2})$ into the same general form (\ref{b4}) as for
$V^{\text{SM}}(q^{2})$ so that the MSSM form factors can also be obtained
from (\ref{b5})-(\ref{b7}).

For the LL MSSM penguin we find, after neglecting terms of order 
$m^{2}_{q}/m^{2}_{\tilde{g}}$ and $m^{2}_{b}/m^{2}_{\tilde{g}}$,
\widetext
\begin{eqnarray}
\label{b22}
\alpha^{\text{MSSM}}(q^{2})&=&\sum_{j}\;\Lambda^{bq}_{LLj} \int^{1}_{0}
dx \int^{1-x}_{0} dy 
\{C_{2}(G) \frac{2xy+4y(y-1)}{Z_{j}(x,y)} \nonumber  \\
&& +[-C_{2}(G) +2C_{2}(R)]  
 \frac{2xy-2y(1-2y)}{Z^{\prime}_{j}(x,y)}\}, \\
\label{b23}
\beta^{\text{MSSM}}(q^{2})&=&\sum_{j}\;\Lambda^{bq}_{LLj}\int^{1}_{0}
dx \int^{1-x}_{0}dy   
   \{C_{2}(G)\frac{2x^{2}-2x+6xy+4y(y-1)}{Z_{j}(x,y)} \nonumber \\
&&    +[-C_{2}(G)+2C_{2}(R)]  
    \frac{2x^{2}-4x+6xy+2(y-1)(2y-1)}{Z^{\prime}_{j}(x,y)} \}  
\end{eqnarray}
and
\begin{eqnarray}
\label{b24}
F^{L}_{2}(q^{2})&=&-m_{q}\sum_{j}\;\Lambda^{bq}_{LLj}\int^{1}_{0} dx
\int^{1-x}_{0} dy   
\{C_{2}(G) \frac{xy}{Z_{j}(x,y)}  \nonumber  \\
&& +[-C_{2}(G)+2C_{2}(R)] \frac{xy}{Z^{\prime}_{j}(x,y)}\}, \\
\label{b25}
F^{R}_{2}(q^{2})&=&m_{b}\sum_{j}\;\Lambda^{bq}_{LLj}\int^{1}_{0} dx
\int^{1-x}_{0} dy   
   \{C_{2}(G)\frac{x^{2}+x(y-1)}{Z_{j}(x,y)}  \nonumber  \\
&&        +[-C_{2}(G)+2C_{2}(R)]\frac{x^{2}+x(y-1)}{Z^{\prime}_{j}(x,y)}\}
\end{eqnarray}
\narrowtext
where
\begin{eqnarray}
\label{b26}
Z_{j}(x,y)&=&1-x+\tilde{x}_{j}+q^{2}[xy+y(y-1)]/m^{2}_{\tilde{g}}, \\
\label{b27}
Z^{\prime}_{j}(x,y)&=&x+\tilde{x}_{j}(1-x)+q^{2}[xy+y(y-1)]/m^{2}_{\tilde{g}}
\end{eqnarray}
with $\tilde{x}_{j}\equiv m^{2}_{\tilde{d}_{L_{j}}}/m^{2}_{\tilde{g}}$.

As $q^{2} \ll m^{2}_{\tilde{d}_{L_{j}}}$ we can set $q^{2}=0$ in (\ref{b26}) 
and (\ref{b27}) to get the LL penguin contributions
\begin{eqnarray}
\label{b28}
F^{L}_{1}(0)&=&\sum_{j}\;\Lambda^{bq}_{LLj}[C_{2}(G)A(\tilde{x}_{j})
                        +C_{2}(R)B(\tilde{x}_{j})], \\
\label{b29}
F^{R}_{1}(0)&=&0,   \\
\label{b30}
\frac{1}{m_{q}}F^{L}_{2}(0)&=&\frac{1}{m_{b}}F^{R}_{2}  \nonumber  \\
  &=& \sum_{j}\;\Lambda^{bq}_{LLj}[C_{2}(G)C(\tilde{x}_{j})-
      C_{2}(R)D(\tilde{x}_{j})]
\end{eqnarray}
where
\begin{eqnarray}
\label{b31}
A(x)&=&\frac{1}{6(1-x)^{4}}[3-9x+9x^{2}-3x^{3}  \nonumber  \\
        && +(1-3x^{2}+2x^{3})\ln x],  \\
\label{b32}
B(x)&=&\frac{-1}{18(1-x)^{4}}[11-18x+9x^{2}-2x^{3}+6 \ln x],  \\
\label{b33}
C(x)&=&\frac{-1}{4(1-x)^{3}}[1-x^{2} +2x \ln x],  \\
\label{b34}
D(x)&=&\frac{-1}{6(1-x)^{4}}[2+3x-6x^{2}+x^{3}+6x \ln x].
\end{eqnarray}

For the LR MSSM penguin
\begin{eqnarray}
\label{b35}
\alpha^{\text{MSSM}}(q^{2})&=&-\frac{m_{\tilde{g}}}{m_{q}}\sum_{j}\;
\Lambda^{bq}_{LRj}\int^{1}_{0} dx \int^{1-x}_{0} dy  \nonumber  \\
         && \times \{C_{2}(G)\frac{x+2y-1}{Z_{j}(x,y)}   \nonumber \\
         &&+[-C_{2}(G)+2C_{2}(R)]\frac{x+2y-1}{Z^{\prime}_{j}(x,y)}\}\\
\label{b36}
\beta^{\text{MSSM}}(q^{2})&=&0
\end{eqnarray}
and
\begin{eqnarray}
\label{b37}
F^{L}_{2}(q^{2})&=&m_{\tilde{g}}\sum_{j}\;\Lambda^{bq}_{LRj}\int^{1}_{0} dx
\int^{1-x}_{0} dy \{C_{2}(G)\frac{x-1}{Z_{j}(x,y)}  \nonumber  \\
        &&+[-C_{2}(G)+2 C_{2}(R)]\frac{x}{Z^{\prime}_{j}(x,y)}\}, \\
\label{b38}
F^{R}_{2}(q^{2})&=&0.
\end{eqnarray}
Again we can set $q^{2}=0$ to obtain for the LR penguin contributions
\begin{equation}
\label{b39}
F^{L}_{1}(0)=F^{R}_{1}(0)=F^{R}_{2}(0)=0
\end{equation}
and
\begin{equation}
\label{b40}
F^{L}_{2}(0)=m_{\tilde{g}}\sum_{j}\;\Lambda^{bq}_{LRj}[C_{2}(G)
E(\tilde{x}_{j})-4 C_{2}(R)C(\tilde{x}_{j})]
\end{equation}
with
\begin{equation}
\label{b41}
E(x)=\frac{-1}{(1-x)^{2}}[1-x+x\ln x] .
\end{equation}

The RR and RL penguins are obtained from the above by the replacements
$\Lambda_{LLj}\rightarrow \Lambda_{RRj}$ and $\Lambda_{LRj} \rightarrow 
\Lambda_{RLj}$ respectively together with $(-m_{q}\alpha^{\text{MSSM}}) 
\leftrightarrow (m_{b}\beta^{\text{MSSM}})$ and $F^{L}_{(1,2)} 
\leftrightarrow F^{R}_{(1,2)}$.

The total $q^{2}=0$ MSSM form factors are therefore
\begin{eqnarray}
\label{b42}
F^{L}_{1}(0)&=&\sum_{j}\;\Lambda^{bq}_{LLj}[C_{2}(G)A(\tilde{x}_{j})
                +C_{2}(R)B(\tilde{x}_{j})],   \\
\label{b43}
F^{R}_{1}(0)&=&\sum_{j}\;\Lambda^{bq}_{RRj}[C_{2}(G)A(\tilde{x}_{j})
                +C_{2}(R)B(\tilde{x}_{j})],  \\
\label{b44}
F^{L}_{2}(0)&=&\sum_{j}\;\{[m_{q}\;\Lambda^{bq}_{LLj}
+m_{b}\;\Lambda^{bq}_{RRj}]  \nonumber  \\
      && \times [C_{2}(G)C(\tilde{x}_{j})-C_{2}(R)D(\tilde{x}_{j})] \nonumber  \\
      && +m_{\tilde{g}}\;\Lambda^{bq}_{RLj}[C_{2}(G)E(\tilde{x}_{j})
                -4C_{2}(R)C(\tilde{x}_{j})]\},   \\
\label{b45}
F^{R}_{2}(0)&=&\sum_{j}\{[m_{b}\;\Lambda^{bq}_{LLj}+
   m_{q}\;\Lambda^{bq}_{RRj}]   \nonumber  \\
    &&  \times [C_{2}(G)C(\tilde{x}_{j})-C_{2}(R)D(\tilde{x}_{j})] \nonumber \\
    && +m_{\tilde{g}}\;\Lambda^{bq}_{LRj}[C_{2}(G)E(\tilde{x}_{j})
                -4C_{2}(R)C(\tilde{x}_{j})]\}.
\end{eqnarray}
The results for the $F^{(L,R)}_{2}(0)$ MSSM form factors agree with those of 
\cite{Gerard84}. However for the $F^{(L,R)}_{1}(0)$ form factors, whereas our
$C_{2}(R)$ term is the same as that of \cite{Gerard84} and \cite{Bert87}, the
$A(x)$ function occurring in the $C_{2}(G)$ term differs from that of
\cite{Gerard84} by $-(1-x)^{-2}\ln x$ and bears little resemblance to the
$F(x)$ function of \cite{Bert87}. Note though that our result for $C_{2}(G) 
A(x)+C_{2}(R)B(x)$ is the same as the function $P_{F}-\frac{1}{9}P_{B}$ given 
in \cite{Barb97}.

The MSSM calculations are described in \cite{Abel96}. Two-loop MSSM RGEs
were used for the gauge and Yukawa couplings and one-loop MSSM RGEs for the
other SUSY parameters. Full flavour dependence was included in the running,
with one-loop QCD and stop/gluino corrections to the physical top mass 
from \cite{Bag}. 
The unification scale boundary conditions were a 
universal scalar mass $m_{0}$, universal gaugino mass $m_{1/2}$ and a 
universal soft SUSY-breaking trilinear scalar coupling $A$.
After minimisation at the scale $m_{t}$ of the full one-loop Higgs effective 
potential, which included all contributions from the matter and gauge 
sectors, we are left with a four-dimensional parameter space $\{m_{0},
m_{1/2},A, \tan \beta\}$, where $\tan \beta \equiv v_{2}/v_{1}$ is the ratio 
of the vacuum expectation values of the two Higgs fields, together with
the sign of the coupling $\mu $ between the two Higgs fields.  
The physical Higgs masses were determined using the approximation
to the RG-improved Higgs masses described in \cite{Haber}.
Mass eigenvalues and diagonalisation matrices for the $d$ squarks were 
generated for a selection of data sets in the parameter space 
$150 \leq m_{0} \leq 1150;\;150 \leq m_{1/2} 
\leq 1150;\;150 \leq |A| \leq 1150$ (units of GeV) and 
$ 2 \leq \tan \beta \leq 48$  which satisfied
current experimental constraints (see \cite{exptrev}), and yielded a 
neutralino as the lightest supersymmetric particle. 
We also imposed the condition that the Standard Model like 
minimum be the global one as has become customary~\cite{ccb1}. 
However it should be noted that, as pointed out in \cite{Abel98},
this traditional condition is not {\em sufficient} to avoid cosmological 
problems. For this one should employ the slightly 
more restrictive condition in \cite{Abel98}. 

The allowed values of $A$ become more restricted 
by unphysical (charge and colour breaking) minima as 
$\tan \beta $ increases from its fixed point value of $\tan\beta 
\approx 1.5$~\cite{Abel98}.
The avoidance of unphysical minima gives a bound of 
$m_0/m_{1/2} \gtrsim 1$ at the low fixed point which
drops away to about 0.4 at intermediate values of $\tan\beta$. 
However the minimum bound on $m_0/m_{1/2}$ is for $A\sim m_0$ and it 
increases 
quadratically in $A$ away from this value~\cite{Abel98}, 
so that effectively $0.5 < A <1.5 m_0$ 
at intermediate $\tan\beta $ values.
Data sets for negative $A$ were therefore more restricted 
in this region regardless of the sign 
of $\mu $, with all but those near $m_{0}$ 
producing colour breaking minima.
Near the high $\tan\beta $ fixed point, where the bottom Yukawa coupling 
is large, the analysis of \cite{Abel98}
is no longer valid and the parameter space becomes once again less 
restricted here. Negative and quite large values of $A$ are allowed 
(and even favoured) over positive ones in this region. 

Finally we should add that additional and probably very restrictive 
constraints on $m_0$ especially at low $\tan\beta$ come from the need 
to avoid neutralino dark matter overclosing the universe. This was examined 
recently in \cite{ellis97} but has not been included in our analysis here. 

The magnitudes of the MSSM form factors satisfy 
$|F^{R}_{2}| > |F^{L}_{1}| \gtrsim |F^{L}_{2}| \gg |F^{R}_{1}|$ 
for all regions of the allowed parameter space apart from the narrow 
region $\tan \beta =2,\;m_{1/2}=150$ and $m_{0} \gtrsim 1000 $ where
$|F^{L}_{1}|$ is slightly smaller than $|F^{R}_{2}|$. 
Outside this region the ratio $|F^{R}_{2}|/|F^{L}_{1}| $ exceeds unity and
increases strongly with $\tan \beta $.  
For $\tan \beta =2$, the ratio ranges from $\approx 2 $ for $m_{1/2}=250$ 
and $m_{0} \gtrsim 1000 $ to $\approx 9 $ for low $(m_{0},m_{1/2})=(150,250)$. 
For  higher $\tan \beta $, $|F^{R}_{2}|$ becomes more dominant, the 
ratio  increasing to 24--28 for $\tan \beta = 10 $ and 200--225 for
$\tan \beta =48$. The relative sizes of the form factors are due to both 
the mixing coefficients $\Lambda^{bq}_{ABj}$ (\ref{bbb1}) and the functions
$A,\;B,\;C,\;D$ and $E$ of the variable
$\tilde{x}_{j} \equiv m^{2}_{\tilde{d}_{j}}/m^{2}_{\tilde{g}}$. If the $j$ 
dependence of $\tilde{x}_{j}$ is neglected, the quantities $\Lambda^{bq}_{AB}
\equiv | \sum_{j}\Lambda^{bq}_{ABj}| $ satisfy $\Lambda^{bq}_{LL} >
\Lambda^{bq}_{RR} > \Lambda^{bq}_{LR} > \Lambda^{bq}_{RL}$ and this accounts 
in the main for the relative sizes of the form factors.
The large values of the form factors at high $\tan \beta $ are 
due to an interplay of two factors; (i) the light gluino mass 
($m_{\tilde{g}} \approx 360$ GeV) associated with $m_{1/2}=150$ and (ii) a
gluino mass lying within the range of $\tilde{d}$ masses such that the 
variable $\tilde{x}_{j}$  is close to unity for several values of $j$.

The result that $F^{R}_{2}$ is the largest MSSM form factor indicates that, 
in contrast to the SM, the magnetic dipole transition dominates the $b$ decay
process in the MSSM. To compare with the SM, we note that the ratio of 
the largest MSSM and SM form factors is
$|F^{R}_{2}(\mathrm{MSSM})|/|F^{L}_{1}(\mathrm{SM})(q^{2}=0)| \leq 0.4$ GeV.

The phases of the MSSM form factors change very little over the allowed
parameter space. The phases of $F^{L}_{1}$ and $F^{(L,R)}_{2}$
are independent of the sign of $A$ and, for $\mu < 0$, are 
approximately equal at $\approx -2.8$ for $b \rightarrow d$ and 
$\approx -0.016$ for $b \rightarrow s$. For $\mu > 0$ the phases of 
$F^{(L,R)}_{2}$ are shifted by $\pi $. The phase of $F^{R}_{1}$ varies a 
little with $m_{0}$ and $m_{1/2}$ and depends on the sign of $A$, being
approximately that of $F^{L}_{1}$ for $A > 0$ and shifted by $\pi $ from
that of $F^{L}_{1}$ for $A < 0$.
These MSSM phases for $\mu < 0$ 
are comparable to the corresponding SM phases so that the 
magnitude of the phase difference between the dominant 
MSSM form factor $F^{R}_{2}$ and that of the SM form factor 
$F^{L}_{1}(q^{2}=0)$ is $\approx 0.4$ for 
$b \rightarrow d $ and $\approx 0.01$ for $b \rightarrow s $ for $\mu < 0$ 
and $\approx 2.7 $ and $\approx 3.1$ for $\mu > 0$. Hence, after allowance 
for the negative sign in (\ref{bbb1}), we conclude that the superpenguins
and ordinary penguins have the same sign for $\mu > 0$ and opposite sign
for $\mu < 0$.

\section{SUSY effects in \lowercase{
$b \rightarrow q q^{\prime} \bar{q}^{\prime}$}}

One albeit crude measure of the effects of SUSY in the decays $b \rightarrow
q q^{\prime} \bar{q}^{\prime}$ is the relative size of the integrated 
decay rates for the MSSM, taken in isolation, and for the SM. 
In computing these decay rates from (\ref{b48})
we impose the low $q^{2}$ cutoff $q^{2} \geq 1$ GeV to avoid non-perturbative
long distance effects.

The largest effects of SUSY on the decay rates occur for high 
$\tan \beta $ and low $(m_{0},m_{1/2}) $. For $\tan \beta = 48$ and 
$A=-300$ the ratio 
$\Gamma^{\mathrm{Peng}}(\mathrm{MSSM})/\Gamma^{\mathrm{Peng}}(\mathrm{SM})$ 
has a maximum value at $(m_{0},m_{1/2})=(275,150)$ of $\approx 0.10 
\;(b \rightarrow d)$ and $\approx 0.085 \;(b \rightarrow s)$ 
for $\mu > 0$ and $\approx 0.09 \;(b \rightarrow d)$ and $\approx 0.08 
\;(b \rightarrow s)$ for $\mu < 0$. The ratio exceeds $10^{-2}$ at 
$\tan \beta =48$ for $(m_{0},m_{1/2})$ ranging from $(225,150)$ to 
$(275,225)$. 
However, for lower $\tan \beta $ this ratio has a much smaller maximum 
value; for $\tan \beta = 10$ it is $1 \times 10^{-4}$ at 
$(m_{0},m_{1/2})=(275,225)$ and $3 \times 10^{-4}$ at $(550,150)$ for 
$\tan \beta =2$. The ratio decreases rapidly for large values of 
$m_{1/2}$ due mainly to the increase of the gluino mass in (\ref{bbb1}) 
from $\approx 360$ at $m_{1/2}=150$ to $\approx 1875 $ at $m_{1/2}=850$.
These findings for low and medium $\tan \beta $ differ from the 
earlier estimates \cite{Bert87} that the 
SUSY and SM contributions to $\Gamma(b \rightarrow s q^{\prime}
\bar{q}^{\prime})$ were of comparable size. However, these early estimates 
were based on the assumption that $b \rightarrow s q^{\prime} 
\bar{q}^{\prime}$ could be described solely by the LL penguin form factors
$F^{(L,R)}_{1}(0)$ and our study shows $F^{R}_{2}$ to be the dominant form
factor.

In SUSY there is also a contribution to $b \rightarrow q q^{\prime} 
\bar{q}^{\prime}$ from the box diagrams of Fig. \ref{fig8} 
for which the amplitude is \cite{Bert87,Inami81}
\widetext
\begin{eqnarray}
\label{b53}
M^{\text{Box}}&=& \frac{i g^{2}_{s}}{4 \pi^{2}}\;
[\bar{u}_{q}(p_{q})T^{a}\gamma_{\mu}P_{L}u_{b}(p_{b})]  
\; \{J^{1}_{LL}\;[\bar{u}_{q^{\prime}}(p_{q^{\prime}})
\gamma^{\mu}T^{a}P_{L}v_{\bar{q}^{\prime}}(p_{\bar{q}^{\prime}})] 
\nonumber  \\
&& +J^{2}_{LR}[(\bar{u}_{q^{\prime}}(p_{q^{\prime}})
\gamma^{\mu}T^{a}P_{R}v_{\bar{q}^{\prime}}(p_{\bar{q}^{\prime}})]\} 
 +[\bar{u}_{q}(p_{q})\gamma_{\mu}P_{L}u_{b}(p_{b})]  \nonumber  \\
&& \times
 \{J^{3}_{LL}\;[\bar{u}_{q^{\prime}}(p_{q^{\prime}})
\gamma^{\mu}P_{L}v_{\bar{q}^{\prime}}(p_{\bar{q}^{\prime}})] 
+J^{3}_{LR}[\bar{u}_{q^{\prime}}(p_{q^{\prime}})
\gamma^{\mu}P_{R}v_{\bar{q}^{\prime}}(p_{\bar{q}^{\prime}})]\} 
+ (L \leftrightarrow R)
\end{eqnarray}
\narrowtext
where
\begin{equation}
\label{b54}
J^{\alpha}_{AB} \equiv \sum_{\tilde{d}_{j}}\;\sum_{\tilde{q}_{i}}\;J_{\alpha}
(\tilde{x}_{j},\tilde{y}_{i})\;\Lambda^{bq}_{AAj}\;
V^{iq^{\prime}\;*}_{\tilde{q}B}\;V^{q^{\prime}i}_{\tilde{q}B}
\end{equation}
with
\begin{eqnarray}
\label{b55}
J_{1}& = & \frac{7}{6}\;g(\tilde{x}_{j},\tilde{y}_{i}) 
-\frac{2}{3}\;f(\tilde{x}_{j},\tilde{y}_{i}),   \\
J_{2}& = &- \frac{1}{3}\;g(\tilde{x}_{j},\tilde{y}_{i}) 
+\frac{7}{3}\;f(\tilde{x}_{j},\tilde{y}_{i}),   \\
J_{3}& = & \frac{2}{9}\;g(\tilde{x}_{j},\tilde{y}_{i}) 
+\frac{4}{9}\;f(\tilde{x}_{j},\tilde{y}_{i}).
\end{eqnarray}
Here $\tilde{y}_{i} \equiv m^{2}_{\tilde{q}_{i}}/m^{2}_{\tilde{g}}$, the 
squark $\tilde{q}$ is $\tilde{u}$ for $q^{\prime}=u$ and $\tilde{d}$ for
$q^{\prime}=d$ and the box functions are \cite{Inami81}
\begin{eqnarray}
\label{b56}
g(x,y) &=&\frac{1}{x-y}[\frac{1}{y-1}+(\frac{x}{x-1})^{2}\ln x
-(x \rightarrow y)]  \\
f(x,y) &=&\frac{1}{y-x}[\frac{1}{y-1}+\frac{x}{(x-1)^{2}}\ln x
-(x \rightarrow y)].  
\end{eqnarray}

For the allowed regions of parameter space the box amplitudes satisfy
$|J^{2}_{LR}| > 3 (|J^{1}_{LL}|,\\ |J^{3}_{LL}|,|J^{3}_{LR}|) \gg 
(|J^{2}_{RL}|,|J^{1}_{RR}|,|J^{3}_{RL}|,|J^{3}_{RR}|)$ apart from in 
the region $\tan \beta =2,\;m_{1/2}=150$ and $m_{0}> 650$ where
$|J^{1}_{LL}|$ becomes slightly larger than $|J^{2}_{LR}|$. 
The four largest box amplitudes
are generally of the same order as the MSSM penguin amplitudes $(F^{L}_{1},
F^{(L,R)}_{2})$; the remaining four are negligible, being smaller by a 
factor of at least $10^{5}$ and comparable to the MSSM penguin $F^{R}_{1}$.
For the regions where $\Gamma^{\mathrm{MSSM}}/\Gamma^{\mathrm{SM}} 
\gtrsim 10^{-3}$, the ratio of the largest MSSM box and penguin amplitudes 
$|J^{2}_{LR}|/|F^{R}_{2}|$ is small, varying from only $8 \times 10^{-3}$ 
for the parameters $\tan \beta =48,\;(m_{0},m_{1/2}) = (275,150)$ which
produce the maximum SUSY effects to a maximum of 0.01. 
The ratio does increase to $\approx 0.4 $ for low $\tan \beta \;(=2)$, 
low $m_{1/2}\;(=150)$  and high $m_{0}> 1000$ but
in these regions the SUSY effects are negligible. 
Hence, for the MSSM data sets for which the SUSY penguin effects are 
largest, the SUSY
box amplitudes can be neglected in calculating the decay rates. 

The differential decay rates $d\Gamma^{\mathrm{Peng}}/dq^{2}$ (\ref{b48}) 
for the SM and for the  combined
effects of the SM and the MSSM for the MSSM data set $A=-300,\;\mu > 0,\;
\tan \beta =48,\;(m_{0},m_{1/2})=(275,150)$ which maximises the
SUSY effects show (see Fig. \ref{fig4} and \ref{fig6}) the 
SUSY enhancement of the decay
rates to be significant for most of the range of $q^{2}$ values. The
partial rate CP asymmetries ${\cal A}_{\mathrm{CP}}(q^{2})$, defined 
in (\ref{bb3}), reveal the presence of SUSY for $q^{2} \lesssim 4 m_{c}^{2}$. 
For these values of $q^{2}$ the SM CP asymmetries of $\approx 25 \%$
for $b \rightarrow d $ and $\approx 1.5 \%$ for $b \rightarrow s$ are 
reduced to $\approx 20\%$ and $\approx 1.2\%$ respectively when the 
MSSM contributions are included.

\section{Discussion and conclusions}

We have calculated, from first principles, both the SM and MSSM penguins
that contribute to the decays of the $b$ and $\bar{b}$ quarks. For the 
MSSM in particular there are discrepancies to be found in the 
literature \cite{Gerard84,Bert87}
as to the correct formulae for the $F^{(L,R)}_{1}$ form factors. 
Our results for these form factors differ slightly from those 
of \cite{Gerard84} but our other results agree with \cite{Gerard84}.

\begin{figure}
  \begin{center}
    \leavevmode   
    \hbox{\epsfxsize=0.8\textwidth
      \epsffile{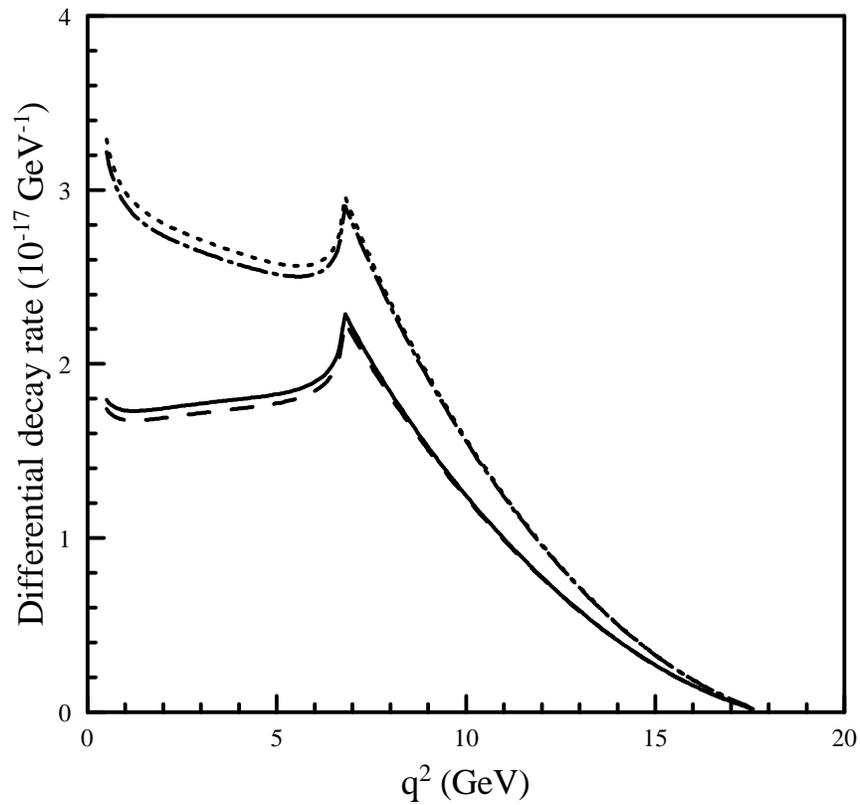}}
  \end{center}
  \caption{Differential SM decay rates for $b \rightarrow s d \bar{d}$
  (solid line) and $ \bar{b} \rightarrow \bar{s} d \bar{d}$ (dashed line). The
  corresponding results for the combined effects of the SM and MSSM are 
  given by the dotted and dot-dashed lines respectively.}
  \label{fig6} 
\end{figure}

\begin{figure}
  \begin{center}
    \leavevmode   
    \hbox{\epsfxsize=0.8\textwidth
      \epsffile{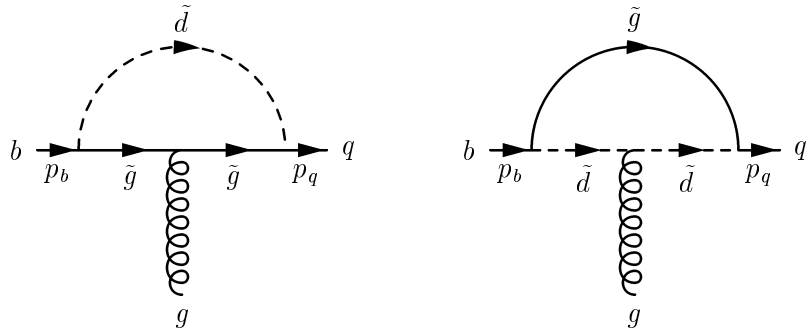}}
  \end{center}
  \caption{MSSM gluino penguins.}
  \label{fig7} 
\end{figure}

\begin{figure}
  \begin{center}
    \leavevmode   
    \hbox{\epsfxsize=0.8\textwidth
      \epsffile{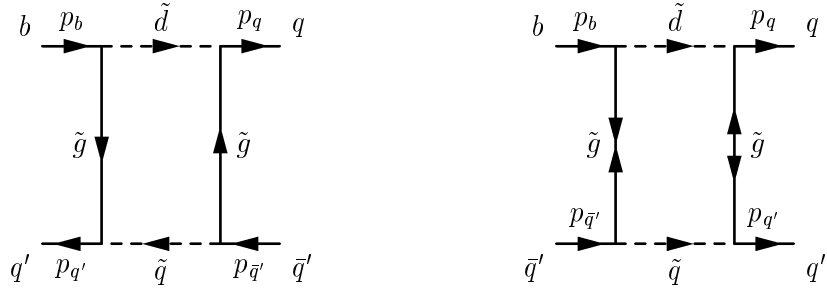}}
  \end{center}
  \caption{MSSM box contributions to $b \rightarrow q q^{\prime}
  \bar{q}^{\prime}$.}
  \label{fig8} 
\end{figure}

Because of the presence of strong phases in the contributions to the 
SM penguin amplitudes from $u$ and $c$ quarks we find that the decay rates
for $b \rightarrow d q^{\prime} \bar{q}^{\prime}$ is significantly
different to the rate $ \bar{b} \rightarrow \bar{d} q^{\prime} 
\bar{q}^{\prime}$ even for quarks in isolation (see Fig. \ref{fig6}).

The SUSY enhancement of the gluon-mediated exclusive hadronic $b$ decays
within the constrained MSSM model can be at the several percent level in
certain regions of the $(A,\;\tan \beta ,\;m_{0},\;m_{1/2})$ parameter
space. In these regions the SUSY penguin processes dominate the SUSY box
processes with the consequence that the $b$ decays in the MSSM 
are driven by the magnetic dipole transition
rather than the electric dipole transition of the SM.

QCD corrections arising from renormalisation of the present short distance
results down from the electroweak scale to the scale $m_{b}$ are not 
likely to alter the finding that the magnetic
amplitude is dominant in the MSSM as the QCD induced mixing effects
\cite{Buch96,Ali97} produce an enhancement of the magnetic dipole operators
in the $\Delta B =1$ effective Hamiltonian relative to the current-current
penguin operators associated with the electric dipole amplitude.
Furthermore, G\'{e}rard and Hou \cite{Gerard91} have noted that the result
(\ref{b19}) for the SM form factor $F^{L}_{1}(q^{2})$ already contains the
dominant part of the QCD corrections for the current-current penguin
operators and, therefore, that the main effects of QCD corrections will be
the renormalisation of the strong coupling constant from $\alpha_{s}(M_{W})$
to $\alpha_{s}(m_{b})$. This would have the effect of increasing 
the penguin decay rates of the SM by the factor 
$\eta \equiv \alpha_{s}(m_{b})/\alpha_{s}(M_{W})\approx 1.84$ and also 
increasing the MSSM penguin amplitudes relative to those of the SM. 

Detection of new physics in the hadronic decay amplitudes of the $b$ quark
through a study of deviations from the predictions of the SM in the 
patterns of CP violation in $B_{d}$ decays is complicated, on the one 
hand by the interplay
between the cumulative effects in the SM of the $q^{2}$-dependent strong
phases in $F^{L}_{1}$ and the weak CKM phases from the contributing $u,\;c$
and $t$ quarks and, on the other hand, by MSSM phases comparable in  
magnitude to the SM weak phases and which can give  constructive or 
destructive interference depending upon the details of the soft 
SUSY-breaking mechanism.

\acknowledgments
W N Cottingham wishes to thank the PPARC theory travel fund for support 
for travel to James Cook University.

\newpage

\newpage

\newpage

\newpage

\newpage

\newpage

\newpage

\newpage

\end{document}